\documentclass[aps,pra,reprint,groupedaddress,showpacs]{revtex4-1}

\usepackage{graphicx}
\usepackage{dcolumn}
\usepackage{bm}
\usepackage{amssymb}
\usepackage{mathrsfs}

\begin{document}

\title{Experimental study of a phase-sensitive heterodyne detector}


\author{Heng Fan}
\author{Dechao He}
\author{Sheng Feng}
\email[]{fengsf2a@hust.edu.cn}
\affiliation{
MOE Key Laboratory of Fundamental Quantities Measurement, School of Physics, Huazhong University of Science and Technology, Wuhan 430074, China}


\date{\today}

\begin{abstract}
It is believed that the quantum behaviors of homodyne detectors and traditional heterodyne detectors can be fully understood in the context of the quantum theory of optical detection. According to the theory, a 3 dB extra quantum noise has been predicted in a traditional heterodyne detector, as a phase-insensitive device, due to the existence of the image sideband vacuum. However, regarding the noise performance of a phase-sensitive heterodyne detector, a fundamental dilemma inevitably arises: On one hand, the detector should suffer the 3 dB noise penalty caused by the image sideband vacuum, on the other hand, it, as a phase-sensitive device, should be noise free at the quantum level. We report on an experiment on the quantum noise performance of a phase-sensitive heterodyne detector with a bichromatic local oscillator. The results show that the studied detector is noise free, i.e., the quantum noise of the image sideband vacuum is absent in the observation. Revealing the mechanism for the absence of the image vacuum noise will be important for a full understanding of the origin of the quantum noise in optical detection.
\end{abstract}


\maketitle


\section{introduction}
The quantum theory of optical detection \cite{personick1971,yuen1978,shapiro1979,yuen1980,yuen1983,schumaker1984,yamamoto1986,collett1987,carmichael1987,ou1987,caves1994,ou1995,haus2011}, developed in consistency with the uncertainty principle, is commonly considered as a correct and complete description of the detection characteristics of non-classical light. However, our confidence about the perfection of the detection theory may be shaken by the new achievements in the study of the uncertainty principle: Heisenberg's measurement uncertainty relation may not hold in full generality \cite{ozawa2003,erhart2012,rozema2012,branciard2013,lorenzo2013,weston2013,baek2013,ringbauer2014,kaneda2014}. If this turns out to be true, doubt may be cast onto a famous prediction, made by the detection theory in accordance with the measurement uncertainty relation, that a 3 dB extra quantum noise takes place in joint measurement of conjugate quadratures of light field with heterodyne detectors \cite{haus1962a,yuen1980,yamamoto1986,caves1994,buonanno2003}. Bearing the above understanding in mind, we study a phase-sensitive heterodyne detector, with which the quantum theory of optical detection is put under new experimental test concerning the origin of the quantum noise of optical detectors.

As is well known, the quantum theory of detection deals with optical heterodyning essentially relying on the concept of image sideband vacuum mode \cite{personick1971,yuen1978,shapiro1979,yuen1980,yuen1983,yamamoto1986,collett1987,caves1994,haus2011}. A traditional heterodyne detector is a phase-insensitive device and suffers a 3 dB noise penalty caused by the extra quantum noise introduced by the image sideband vacuum mode (Table \ref{tab:dilemma}) \cite{personick1971,yuen1980,yuen1983,yamamoto1986,caves1994}. If the image sideband vacuum mode is also excited into a coherent state to the same level as the signal mode, the heterodyne detector becomes phase sensitive and is free of the noise penalty \cite{collett1987,caves1982}. 

\begin{table}
\caption{\label{tab:dilemma} Optical detectors and their quantum noise performances. BLO stands for bichromatic local oscillator. The ``{\small$\surd$}'' marks mean that ``the detector is phase sensitive'' or ``image band vacuum is involved in the detection'', and ``{$\times$}'' means the opposite of ``{\small$\surd$}''. The question mark ``?'' represents the theoretical dilemma faced by the existing quantum theory of optical detection, regarding the quantum noise of a BLO heterodyne detector.
}
\begin{ruledtabular}\vspace{0.05in}
\begin{tabular}{c|c|c|c}
\ \ \ \  \ \ \ & \ \ \ \ &  traditional \ \ \ \ & BLO  \\ 
\ \ \ \  \ \ \ & homodyne\ \ \ \ &  heterodyne\ \ \ \ & heterodyne \\ 
\ \ \ \  \ \ \ & detector\ \ \ \ &  detector\ \ \ \ & detector \\ \hline
phase &  \ \ \ \ &  \ \ \ \ &  \ \ \  \ \ \\
sensitivity &  $\surd$ \ \ \ \ & \Large{$\times$} \ \ \ \ & $\surd$ \ \  \ \ \\
 &   \ \ \ \ &  \ \ \ \ &  \ \  \ \ \\ \hline
image &  \ \ \ \ &  \ \ \ \ & \ \  \ \ \ \\
sideband &  \Large{$\times$} \ \ \ \ & \mbox{$\surd$} \ \ \ \ & \mbox{$\surd$}\ \  \ \ \\ 
vacuum &  \ \ \ \ &  \ \ \ \ & \ \  \ \ \\ \hline
quantum &  \ \ \ \ & \ \ \ \ &  \ \  \ \ \\
noise  &  noise free \ \ \ \ & 3 dB penalty \ \ \ \ & \Large{?}  \ \ \\
performance &   \ \ \ \ &  \ \ \ \ &  \  \ \ \ \\
\end{tabular}
\end{ruledtabular}
\end{table}

Notwithstanding, it is nontrivial to describe the quantum behavior of a phase-sensitive heterodyne detector with a bichromatic local oscillator in the context of the current detection theory: (Table \ref{tab:dilemma}): On one hand, a 3 dB noise penalty is expected for the detector due to the presence of the image sideband vacuum modes \cite{marino2007}. On the other hand, however, the detector should be noise free at the quantum level because of its phase-sensitive nature \cite{caves1982}. As an indication that a fundamental self inconsistency may reside in the quantum theory of detection, the above dilemma reflects our lack of full understanding of the origin of the quantum noise in optical detection, especially the physics relevant to image sideband vacuum mode. To pursue a self-consistent theoretical development for the studied heterodyne detector, more information is desired from experimental investigation on the quantum noise performance of the detector.

In what follows, we summarize in Sec. II some relevant past works, focusing on how the concept of image sideband vacuum mode was introduced in the detection theory to explain the quantum noise in optical heterodyning. In Sec. III, we extend the theoretical analysis to the case of heterodyne detection with a bichromatic local oscillator. With the orthodox understanding of image sideband vacuum noise, we show that a 3 dB noise penalty should occur in the heterodyne detection, in agreement with a previous work \cite{marino2007} studying heterodyne detection in a similar configuration. Then we show that the heterodyne detector in the studied configuration is indeed phase sensitive and, hence, should be free of noise penalty on the contrary, according to the quantum theory of linear amplifier \cite{caves1982}. With this theoretical dilemma, we report in Sec. IV on an experiment on the noise performance of the phase-sensitive heterodyne detector. The experimental results show that the 3 dB noise penalty was absent in the detector. In Sec. V, we discuss the importance of the experimental observation. A trivial conclusion from the experimental results is that the image sideband vacuum modes contributed no extra quantum noise in the heterodyne detection. To be revealed is the mechanism for the absence of the image sideband vacuum noise in the experimental observation, which should be tightly related to the physics of the origin of the quantum noise in optical detection.


\section{Conventional heterodyne detection}
According to the quantum theory of optical detection, a conventional heterodyne detector, classified as a phase-insensitive device in the theory of linear amplifier \cite{caves1982}, possesses a 3 dB extra quantum noise in comparison with a homodyne detector, caused by the image sideband vacuum mode involved in the detection \cite{personick1971,yuen1978,shapiro1979,yuen1980,yuen1983,yamamoto1986,caves1994}. Although there are different models for theoretically treating the problem of heterodyne detection, most models adopt the imageband-mode concept \cite{personick1971,yuen1978,yuen1980,yuen1983,yamamoto1986,collett1987,caves1994,haus2011} and agree with one another on the 3 dB heterodyne noise penalty.

\begin{figure} 
\includegraphics[scale=0.28]{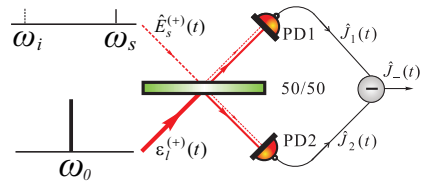}
\caption{\label{fig:het} (color online) Theoretical model for conventional heterodyne detection of an optical signal with a monochromatic local oscillator. The signal light $\hat{E}_s^{(+)}(t)$ interferes with the optical local oscillator $\mathscr{E}_l^{(+)}(t)$ at a balanced (50/50) beamsplitter. The mixed light at each output port of the beamsplitter is collected by a photodetector (PD1 or PD2) and the differenced photocurrent $\hat{J}_-(t)\equiv \hat{J}_1(t)-\hat{J}_2(t)$ is sent to a spectrum analyzer for record.}
\end{figure}

Let consider balanced heterodyne detection, in a traditional configuration (Fig. \ref{fig:het}), of an optical signal at frequency $\omega_s$, $\hat{E}_s^{(+)}(t)=\hat{a}_s e^{-i\omega_s t+i\phi_s}$, where $\hat{a}_s$ is the photon annihilation operator with $\hat{a}_s^\dagger\hat{a}_s$ in units of photons per second. The signal enters the input port of a 50-50 beamsplitter together with a vacuum mode $\hat{E}_i^{(+)}(t)=\hat{a}_i e^{-i\omega_i t+i\phi_i}$ \cite{personick1971,yuen1978,yuen1980,yuen1983,yamamoto1986,collett1987,caves1994,ou1995}. The light entering the signal port of the beamsplitter is combined with a much more powerful coherent light at frequency $\omega_0=\frac{1}{2}(\omega_s+\omega_i)$, $\mathscr{E}_l^{(+)}(t)=\mathscr{E}_le^{-i\omega_0 t+i\phi_0}$. Here the amplitude $\mathscr{E}_l$ is a real number. Both outputs of the beamsplitter are directed onto photodiodes, whose output photocurrent is differenced and then filtered to pick out the beatnote signal at frequency $\Omega\equiv\omega_s-\omega_0$ ($\omega_s>\omega_0$ is assumed for simplicity). Then the result is detection of the quantity $\hat{X}=\hat{a}_se^{-i\phi}+\hat{a}_i^\dagger e^{i\phi}$ \cite{haus1962a,yuen1980,caves1994,yamamoto1986},
\begin{equation} \label{eq:heterojm}
\hat{J}_- = e\mathscr{E}_le^{-i\Omega t+i\delta\phi}\hat{X} +\ \mbox{h.c.}.
\end{equation}
Here $\phi=\phi_0-(\phi_s+\phi_i)/2$, $\delta\phi=(\phi_s-\phi_i)/2$, and $e$ is the charge on the electron. The quantum efficiency is assumed perfect for the moment.

To quantitatively describe the noise penalty in optical detection, one usually makes use of the quantity of noise figure (NF), which is defined as the ratio of the signal-to-noise ratio (SNR) at the input of a detector to that at its output. According to this definition, NF $=$ 3 dB means a 3 dB noise penalty. 

If the detector senses an optical signal in a coherent state, a special kind of quantum state, the SNR at its input is
\begin{eqnarray} \label{eq:snrin}
\mbox{SNR}_{in}=\frac{\bar{N}_\gamma^2}{(\Delta N_\gamma)^2}=\bar{N}_\gamma,
\end{eqnarray}
where $\bar{N}_\gamma$ stands for the average photon number received by the detector and $(\Delta N_\gamma)^2=\bar{N}_\gamma$ is the corresponding photon-number fluctuation for coherent light. The SNR at the detector's output is
\begin{eqnarray} \label{eq:snrout}
\mbox{SNR}_{out}=\frac{P_i}{P_n},
\end{eqnarray}
in which $P_i$ is the average power of the output photoelectric signal and $P_n$ the quantum-noise power of the output signal. One can easily calculate the classical quantity $P_i$ as \cite{oliver1961,haus1962a}
\begin{eqnarray} \label{eq:pi}
P_i=<\left[e \mathscr{E}_l \mbox{Re}(e^{-i\Omega t+i\delta\phi}<\hat{X}>)\right]^2>_s=(e \alpha_s \mathscr{E}_l)^2/2,\ \ 
\end{eqnarray}
wherein $<\cdot>_s$ means statistical average, $\alpha_s=<\hat{a}_s>$, and the load (spectrum analyzer) resistance was, and hereafter, set as one without loss of generality.
The quantum-noise power $P_n$ in a measurement time of one second is
\begin{eqnarray} \label{eq:pn}
P_n&=&<(\Delta \hat{J}_-)^2>_s/2 \nonumber\\
&=&<\left(<\hat{J}_-^2>-<\hat{J}_->^2\right)>_s/2 \nonumber\\
 &=& (e \mathscr{E}_l)^2,
\end{eqnarray}
for input light in a coherent state. Here the factor of 1/2 comes from averaging the power of AC signals over time. If the input light is in a two-mode squeezed state, a quantum correlation may occur between the signal mode and the image sideband mode. In this case, the quantum-noise power $P_n$ becomes \cite{marino2007}
\begin{eqnarray} 
P_n &=& (e \mathscr{E}_l)^2<\left[e^{2s}\cos^2\phi+e^{-2s}\sin^2\phi\right]>_s,
\end{eqnarray}
which agrees with Eq. (\ref{eq:pn}) when the degree of squeezing reduces down to zero, i.e., the squeezing parameter $s=0$. Therefore, the SNR at the output of a conventional heterodyne detector sensing coherent light is, according to Eq. (\ref{eq:snrout}),
\begin{eqnarray} \label{eq:snrout1}
\mbox{SNR}_{out}=\frac{P_i}{P_n}=\frac{(e \alpha_s \mathscr{E}_l)^2/2}{(e \mathscr{E}_l)^2}=\alpha^2_s/2.
\end{eqnarray}
Since $\alpha_s^2$ is the photon number per second, $\alpha^2_s=\bar{N}_\gamma$ for a measurement time of one second. Therefore, the noise figure of the detector is NF $= 10\log_{10} (\mbox{SNR}_{in}/\mbox{SNR}_{out})$ = 3 dB, according to Eqs. (\ref{eq:snrin}) and (\ref{eq:snrout1}).

\begin{figure} 
\includegraphics[scale=0.35]{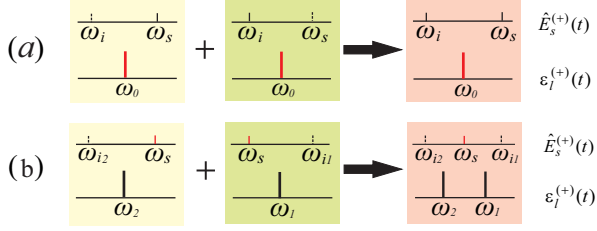}
\caption{\label{fig:pid2psd} (color online) Illustration of two ways to construct a phase-sensitive heterodyne detector out of traditional phase-insensitive heterodyne ones. The center-frequency mode in each case is labeled by red color. (a) A phase-sensitive heterodyne detector with a monochromatic local oscillator. All the frequency modes involved in the detection are excited into coherent states. (b) A phase-sensitive heterodyne detector with a bichromatic local oscillator. Two image sideband modes are in vacuum states in this case.
}
\end{figure}

\begin{figure} 
\includegraphics[scale=0.28]{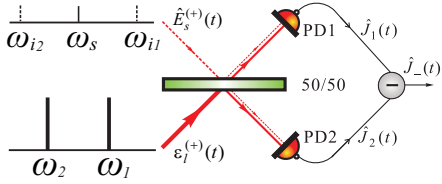}
\caption{\label{fig:hetblo} (color online) Theoretical model for heterodyne detection with a bichromatic local oscillator.}
\end{figure}

The 3 dB degradation of the SNR at the output of a traditional heterodyne detector has been theoretically attributed to the existence of the image sideband vacuum mode $\hat{E}_i^{(+)}(t)$ \cite{personick1971,yuen1978,yuen1980,yuen1983,collett1987,caves1994}, which doubles the quantum-noise power $P_n$ while contributing nothing to the photoelectric signal. However, if the image sideband mode $\hat{E}_i^{(+)}(t)$ is excited into a coherent state to the same power level as the signal mode $\hat{E}_s^{(+)}(t)$ and independently carries another signal, then $P_i$ will be doubled as well at the output of the detector and, hence, NF = 0 dB \cite{collett1987,caves1982}.

\section{phase-sensitive heterodyne detection with a bichromatic local oscillator}
A phase-sensitive heterodyne detector can be constructed out of conventional heterodyne detectors, which are phase-insensitive devices, in two different ways (Fig. \ref{fig:pid2psd}): (1) To excite the image sideband mode into a state similar to the signal mode or (2) to utilize a bichromatic local oscillator instead of a monochromatic local oscillator. As already pointed out above, a phase-sensitive detector in the former case is free of 3 dB noise penalty \cite{caves1982} with no doubt at all, whereas the latter case deserves further investigation.

We consider an optical signal, $\hat{E}_s^{(+)}(t)=\hat{a}_s e^{-i\omega_s t+i\phi_s}$, to be detected by a heterodyne detector with a bichromatic local oscillator $\mathscr{E}_l^{(+)}(t)=(\mathscr{E}_le^{-i\omega_1 t+i\phi_1}+\mathscr{E}_le^{-i\omega_2 t+i\phi_2})/\sqrt{2}$, where $\omega_1-\omega_s=\omega_s-\omega_2=\Omega$ (Fig. \ref{fig:hetblo}). In this case, two image sideband vacuum modes, $\hat{E}_{i1}^{(+)}(t)=\hat{a}_{i1} e^{-i\omega_{i1} t+i\phi_{i1}}$ and $\hat{E}_{i2}^{(+)}(t)=\hat{a}_{i2} e^{-i\omega_{i2} t+i\phi_{i2}}$, are involved in the detection \cite{marino2007}. Obviously, $\omega_{i1}-\omega_1=\omega_2-\omega_{i2}=\Omega$.

Now we show that the NF of a heterodyne detector in this configuration is 3 dB if the image sideband modes introduce extra quantum noise in the detection. The physical quantity detected by the heterodyne detector is $\hat{X}'=\hat{a}_s^\dagger e^{-i\phi_s+i\phi_1}+\hat{a}_se^{i\phi_s-i\phi_2}+\hat{a}_{i1} e^{i\phi_{i1}-i\phi_1}+\hat{a}^\dagger_{i2} e^{-i\phi_{i2}+i\phi_2}$,
\begin{widetext}
\begin{eqnarray} \label{eq:heterojmm}
\hat{J}_- &=& e \mathscr{E}_l^{(+)}(t){\left[\hat{E}_s^{(+)}(t)+\hat{E}_{i1}^{(+)}(t)+\hat{E}_{i2}^{(+)}(t)\right]}^\dagger + e {\left[\mathscr{E}_l^{(+)}(t)\right]}^* \left[\hat{E}_s^{(+)}(t)+\hat{E}_{i1}^{(+)}(t)+\hat{E}_{i2}^{(+)}(t)\right]\nonumber \\
&=& (e\mathscr{E}_l/\sqrt{2})e^{-i\Omega t}\hat{X}'+\ \mbox{h.c.},
\end{eqnarray}
\end{widetext}
where only the terms at the heterodyne frequency $\Omega$ remain in the last step. Accordingly, for an optical signal in a coherent state, the average power of the output photoelectric signal at frequency $\Omega$ is
\begin{eqnarray} \label{eq:pii}
P_i=<\left[\frac{e \mathscr{E}_l}{\sqrt{2}} \mbox{Re}(e^{-i\Omega t}<\hat{X}'>)\right]^2>_s=\frac{(e \alpha_s \mathscr{E}_l)^2}{2}. 
\end{eqnarray}
And the quantum-noise power of the photoelectric signal, similar to Eq. (\ref{eq:pn}), reads
\begin{eqnarray} \label{eq:pnn}
P_n&=&<(\Delta \hat{J}_-)^2>_s/2 \nonumber\\
&=&<\left(<\hat{J}_-^2>-<\hat{J}_->^2\right)>_s/2 \nonumber\\
 &=& (e \mathscr{E}_l/2)^2<\left[4+2\cos(2\Omega t +\phi_2-\phi_1)\right]>_s \nonumber \\
&=& (e \mathscr{E}_l)^2.
\end{eqnarray}
Then, with Eq. (\ref{eq:snrout}), it is straight forward to show SNR$_{out}=\alpha_s^2/2$, the same as Eq. (\ref{eq:snrout1}). Together with Eq. (\ref{eq:snrin}), one can easily show that the NF of the heterodyne detector NF = 3 dB, the nature of a phase-insensitive device \cite{caves1982}, resulted from the image sideband vacuum modes involved in the detection. Similar theoretical results have been obtained previously for such a heterodyne detector that senses light in two-mode squeezed states \cite{marino2007}.

Nonetheless, as we will show in the following, the heterodyne detector with a bichromatic local oscillator is a phase-sensitive device and is supposed to be free of the 3 dB noise penalty, i.e., NF =  0 dB, according to the quantum theory of linear amplifier \cite{caves1982}.

A phase-insensitive device is one, whose average output signal is invariant under arbitrary phase transformations \cite{caves1982}. The output photoelectric signal of the heterodyne detector is, according to Eq. (\ref{eq:heterojmm}),
\begin{eqnarray} \label{eq:psd}
<\hat{J}_->&=& (e\mathscr{E}_l/\sqrt{2})\mbox{Re}\left[e^{i\Omega t}<\hat{X}'>\right]
\nonumber \\
&=&(\sqrt{2}e\mathscr{E}_l\alpha_s\cos\phi')\cos(\Omega t +\delta \phi'),
\end{eqnarray}
where $\phi'=\phi_s-(\phi_1+\phi_2)/2$ and $\delta\phi'=(\phi_2-\phi_1)/2$. The amplitude of $<\hat{J}_->$ is definitely not invariant under phase transformation. For instance, under the transformation $\phi'\rightarrow\phi'+\pi/2$, the amplitude of $<\hat{J}_->$ may drop down to zero from its maximum, thereby showing the phase-sensitive property of the studied heterodyne detector.

Then we come to the theoretical dilemma for the heterodyne detector: On one hand, due to the existence of the image sideband vacuum modes, the detector should suffer a 3 dB noise penalty. On the other hand, as a phase-sensitive device, it should be noise free \cite{caves1982}. At this point, pure theoretical investigation may just lead to fruitless debate and one needs to resort to experimental study for a verdict.

\begin{figure} 
\includegraphics[scale=0.4]{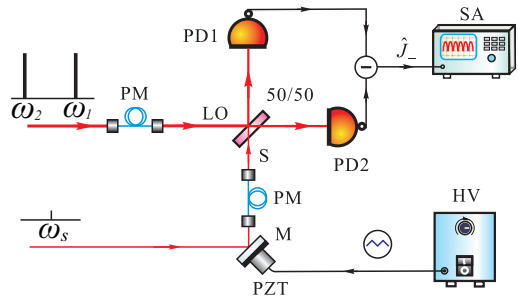}
\caption{\label{fig:setup} (color online) Experimental setup for the study of the quantum noise of a phase-sensitive optical heterodyne detector with a bichromatic local oscillator. PM: Single-mode polarization-maintaining fiber, used as spatial-mode cleaner. M: Mirror. LO: Local oscillator. S: Signal light. 50/50: 50-50 balanced beamsplitter. PD1 \& PD2: Photodiodes. SA: Spectrum analyzer. PZT: Piezoelectric transducer. HV: High-voltage PZT driver.
}
\end{figure}

\section{experimental study of a heterodyne detector with a bichromatic local oscillator}
In this section, we present an experiment on the quantum noise in phase-sensitive heterodyne detection of coherent light with a bichromatic local oscillator. The results of this experiment are important for finding a solution to the potential problem of theoretical self-inconsistency and for one to gain a full understanding of the origin of the quantum noise in optical detection.

The experiment utilized as the light source a laser (Mephisto, Innolight GmbH) emitting a continuous-wave single-frequency coherent light beam (spectral linewidth $<1$ kHz for 0.1 s  measurement time, $\lambda=$ 1064 nm). The laser beam was split into two, each of which was sent through an AOM (Crystal Technology, LLC) for frequency shifting. One of the frequency-shifted beams served as the signal light to be detected, while the other one was used as the bichromatic local oscillator for the detector (Fig. \ref{fig:setup}). Two photodiodes (ETX 500, JDS Uniphase) collected light from the output ports of the 50-50 beamsplitter where optical heterodyning took place and the differenced photocurrent was fed into a spectrum analyzer (Agilent, N9320B) for data record.

\begin{figure} 
\includegraphics[scale=0.3]{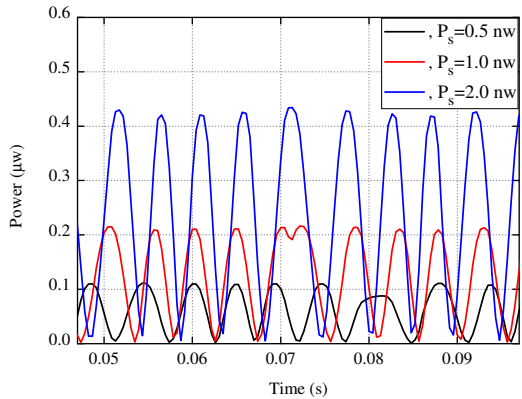}
\caption{\label{fig:psd} (color online) The power of the photoelectric signal at the heterodyne frequency $\Omega=1.3$ MHz, when the relative phase between the signal light and the local oscillator was being scanned. The optical power $P_s$ of the signal light was 0.5$\pm$0.1 nW, 1.0$\pm$0.1 nW, and 2.0$\pm$0.1 nW respectively for the three curves. The power of the bichromatic local oscillator was 2.0 mW for all the three cases. The deviation of the curves from standard cosine functions was primarily due to the residual phase fluctuations between the signal light and the local oscillator.
}
\end{figure}

As is shown by Eq. (\ref{eq:psd}), the heterodyne detector is a phase-sensitive device. To experimentally demonstrate it, one may monitor the output photoelectric signal while scanning the relative phase between the signal beam and the local oscillator. The data presented in Fig. \ref{fig:psd} indeed verify the phase-sensitive property of the heterodyne detector.

In what follows, we demonstrate the capability of the setup to detect the quantum-noise floors of light at appropriate power levels. To this end, we compared the power levels of the detected noises of light with theoretical expectations. The observed noise-power density was -138.7$^{+0.4}_{-0.5}$ dBm/Hz for a 1.0 mW optical oscillator, in which case the theoretical expectation for the quantum-noise power density was -139 dBm/Hz when a 70\% quantum efficiency is taken into account for the detector. Similar result was produced for a 2.0 mW optical oscillator. Moreover, we observed that doubling the power of the optical oscillator resulted in a 2.9$^{+0.5}_{-0.6}$ dB increase, which would be otherwise 6 dB if classical noises of light dominated, in the noise-power spectrum of light. Thereby, we demonstrated that the setup was able to sense the quantum noise of light.

\begin{figure} 
\includegraphics[scale=0.3]{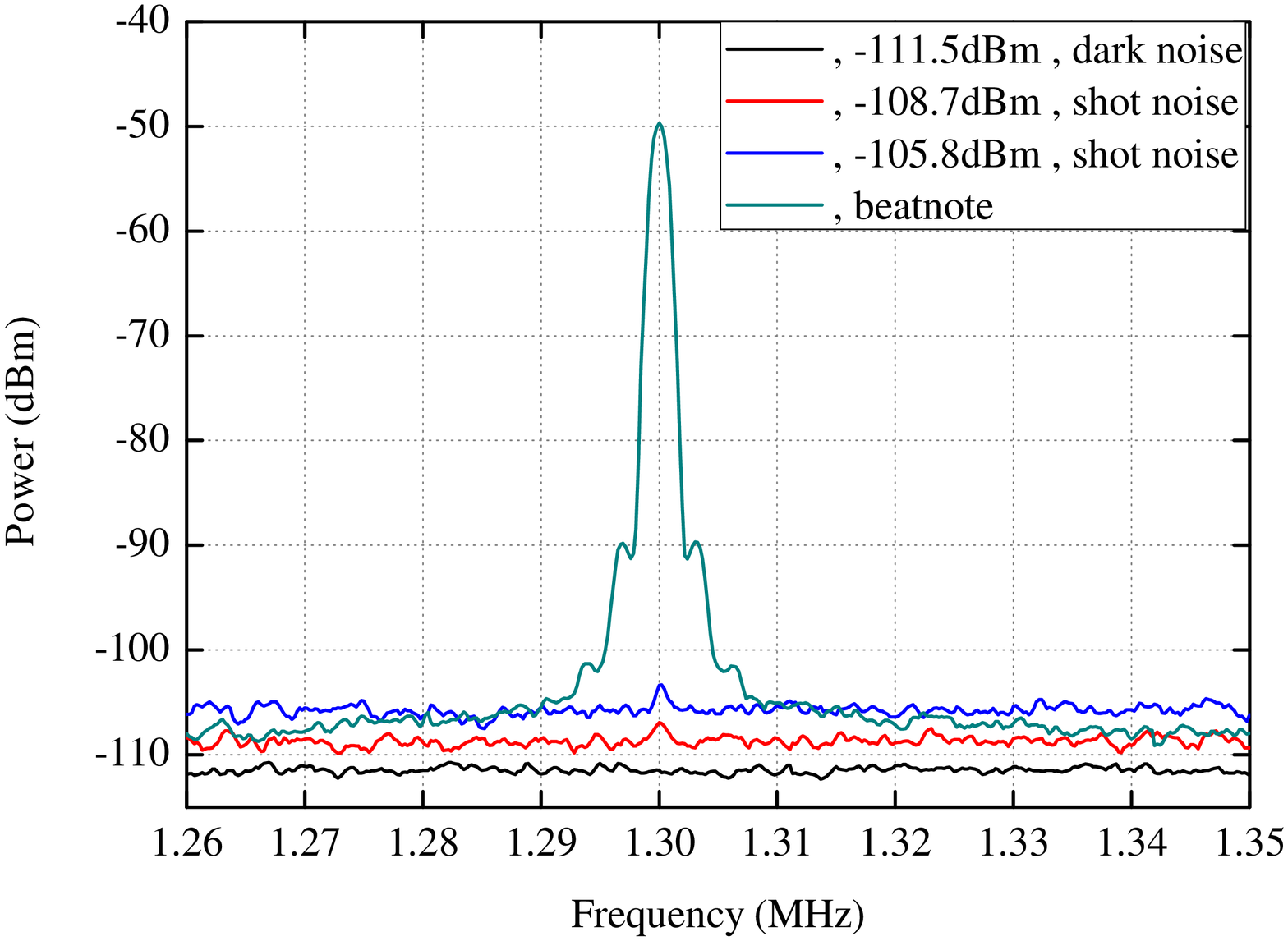}
\caption{\label{fig:data} (color online) A typical (not maximal) heterodyne beatnote signal generated by the heterodyne detector with a bichromatic local oscillator. The optical power of the local oscillator was 1.0 mW. Shown together is the shot-noise curve (red) when the signal light was blocked. The noise floor of the beatnote signal coincided with the shot-noise curve, evidencing that the beatnote was shot-noise-limited. Another shot-noise curve (blue) is present for a 2.0 mW local oscillator. The difference of the power levels of these two curves was 2.9$^{+0.5}_{-0.6}$ dB. Dark noises were excluded in both shot-noise curves. Time of average was 60 s. RBW = 1 kHz.
}
\end{figure}

Now we are ready to present the NF-measurement results for the heterodyne detector. The SNR at the input of the detector was obtained by measuring the optical power of the input signal light, according to Eq. (\ref{eq:snrin}). The output SNR was directly measured with the spectrum analyzer (Agilent, N9320B) by comparing the photoelectric signal power to the quantum-noise power. A typical set of data curves is depicted in Fig. \ref{fig:data} and the final results are shown in Table \ref{tab:nf}.

Obviously, the experimental results show that the phase-sensitive heterodyne detector was noise free \cite{caves1982}, i.e., the 3 dB noise penalty did not occur in the heterodyne detector.

\section{discussions}
The concept of image sideband mode plays a crucial role in the detection theory to describe the quantum nature of optical heterodyne detectors \cite{personick1971,yuen1978,shapiro1979,yuen1980,yuen1983,yamamoto1986,collett1987,caves1994,ou1995,haus2011}. A consensus has been reached on the connection between the quantum noise in heterodyne detection and the image sideband vacuum modes: The image sideband vacuum modes are supposed to introduce extra quantum noise whenever they are involved in optical detection. Now an exception has been found in the heterodyne detector with a bichromatic local oscillator: No extra quantum noise was observed in spite of the existence of the image sideband vacuum modes. This exception indicates our lack of full understanding of the origin of the quantum noise in optical heterodyne detection, especially the physics of the image vacuum noise.

\begin{table}
\caption{\label{tab:nf} Experimentally-determined noise figure (NF) of the heterodyne detector with a 2.0 mW bichromatic local oscillator. SNR$_{in}$ was obtained according to Eq. (\ref{eq:snrin}) with a quantum efficiency of 70\% taken into account for the detector. SNR$_{out}$ was the ratio of the average power of the photoelectric signal to the corresponding shot-noise power, according to Eqs. (\ref{eq:snrout})-(\ref{eq:pn}). Technically, the average power of the photoelectric signal was obtained by averaging the data presented in Fig. \ref{fig:psd}.}
\begin{ruledtabular}
\begin{tabular}{c|c|c|c}
\ \ \ \ $P_{s} $ (nW) \ \ \ & SNR$_{in}$ (dB)\ \ \ \ & SNR$_{out}$ (dB) \ \ \ \ & NF (dB) \\ \hline
0.5$\pm$0.1 & 62.7$\pm$0.9  \ \ \ \ & 63.5$^{+1.0}_{-1.2}$ \ \ \ \ & -0.8$^{+1.2}_{-1.7}$ \\\hline
1.0$\pm$0.1 & 65.7$^{+0.5}_{-0.4}$  \ \ \ \ & 66.5$^{+0.5}_{-0.4}$ \ \ \ \ & -0.8$^{+0.7}_{-0.6}$ \\\hline
2.0$\pm$0.1 & 68.8$^{+0.3}_{-0.2}$  \ \ \ \ & 69.8$\pm$0.4 \ \ \ \ & -1.0$^{+0.5}_{-0.4}$ \\
\end{tabular}
\end{ruledtabular}
\end{table}

Apparently, a trivial prediction in consistency with the experiment can be generated if one assumes in the preceding theoretical calculation that image sideband vacuum modes do not produce extra quantum noise in the phase-sensitive heterodyne detection: The quantum-noise power of the detector at its output, previously described by Eq. (\ref{eq:pnn}), should be
\begin{eqnarray} \label{eq:pnn2}
P_n&=&<(\Delta \hat{J}_-)^2>_s/2 \nonumber\\
 &=& (e \mathscr{E}_l/2)^2<\left[2+2\cos(2\Omega t +\phi_2-\phi_1)\right]>_s \nonumber \\
&=& (e \mathscr{E}_l)^2/2.
\end{eqnarray}
If this is the case, the NF of the phase-sensitive detector would be 0 dB, instead of 3 dB, in good agreement with the experimental observations. 

What is more interesting will come into sight when one considers the NF of the heterodyne detector, provided that the relative phase $\phi=\phi_s-(\phi_1+\phi_2)/2$ between the signal light and the bichromatic oscillator is locked for maximal output signal. In this case, the average power of the output photocurrent is, instead of Eq. (\ref{eq:pii}),
\begin{eqnarray} \label{eq:pii2}
P_i=<\left[\frac{e \mathscr{E}_l}{\sqrt{2}} \mbox{Re}(e^{-i\Omega t}<\hat{X}'>)|_{\phi=k\pi}\right]^2>_s=(e \alpha_s \mathscr{E}_l)^2. \ \ \ \ 
\end{eqnarray}
Here $k$ is an integer. In combination with Eq. (\ref{eq:pnn2}), the SNR at the output of the detector reads
\begin{eqnarray} \label{eq:snrout2}
\mbox{SNR}_{out}=\frac{P_i}{P_n}=\frac{(e \alpha_s \mathscr{E}_l)^2}{(e \mathscr{E}_l)^2/2}=2\alpha^2_s=2\bar{N}_\gamma.
\end{eqnarray}
Then the NF of the detector is NF = -3 dB, which is an incredible result at the first glance! A detector can only degrade, if it does not keep, the SNR of a detected signal. How can the detector increase it? At this point, one may cast doubt on the validity of Eqs. (\ref{eq:pnn2}) and (\ref{eq:snrout2}). Nevertheless, the predicted -3 dB NF for the heterodyne detector for $\phi=k\pi$ is in agreement with the experimental observation.

As a matter of fact, the data presented in Fig. \ref{fig:psd} and Fig. \ref{fig:data} provide one the SNR information of the detector when the relative phase $\phi$ was being scanned. For a 0.5 nW signal light at the input and a 2.0 mW bichromatic local oscillator for the detector, the peak power of the output photoelectric signal was -39.6 dBm (Fig. \ref{fig:psd}) with the noise power of -105.8$^{+0.4}_{-0.5}$ dBm (Fig. \ref{fig:data}). It is not difficult to calculate the SNR of the output signal as SNR$_{out}$ = 66.2$^{+0.4}_{-0.5}$ dB. The SNR of the corresponding input signal was SNR$_{in}$ = 62.7$\pm$0.9 dB (Table \ref{tab:nf}). So the NF of the detector was NF = -3.5$^{+0.9}_{-1.2}$ dB, indicating an SNR increase in the output signal for some chosen values of the relative phase $\phi$ and, thereby, verifying the above paradox about the negative NF for the heterodyne detector. By the way, similar paradox is also found in Eq. (8) of the early work of Haus and Townes \cite{haus1962a} for homodyne detectors. 


The experiment and its agreement with Eq. (\ref{eq:pnn2}) strongly suggest that the image sideband vacuum introduced no extra quantum noise into the phase-sensitive heterodyne detector. To understand the physics of the absence of the image sideband vacuum noise in the observation, one may think of two possibilities: (1) Each of the image sideband vacuum modes contributed no, or negligible, quantum noise, which is inconsistent with the existing theory of heterodyne detection. (2) Both of the image vacuum modes contributed non-negligible quantum noises, but these extra noises were cancelled through some mechanism. If the latter is true, there must be a phase-independent extra term missed in Eq. (\ref{eq:pnn}) and this missing term should be responsible for cancelling the image-mode quantum noise. With this in mind, we indeed can find some phase-independent term in the calculation of the averaged quantum-noise power of the photoelectric signal,
\begin{widetext}
\begin{eqnarray} \label{eq:pnn3}
P_n&=&<\left(<\hat{J}_-^2>-<\hat{J}_->^2\right)>_s/2 \nonumber\\
&=&(e \mathscr{E}_l/2)^2<<\left[e^{-i\Omega t}(\hat{b}_s^\dagger e^{i\phi_1}+\hat{b}_se^{-i\phi_2}+\hat{b}_{i1} e^{-i\phi_1}+\hat{b}^\dagger_{i2} e^{+i\phi_2})+\mbox{h.c.}\right]^2>>_s\nonumber \\
&&-(e \mathscr{E}_l/2)^2<<\left[e^{-i\Omega t}(\hat{b}_s^\dagger e^{i\phi_1}+\hat{b}_se^{-i\phi_2}+\hat{b}_{i1} e^{-i\phi_1}+\hat{b}^\dagger_{i2} e^{+i\phi_2})+\mbox{h.c.}\right]>^2>_s \nonumber \\
&=&(e \mathscr{E}_l/2)^2<2\hat{b}_s\hat{b}_s^\dagger+\hat{b}_{i1}\hat{b}^\dagger_{i1}+\hat{b}_{i2}\hat{b}^\dagger_{i2}+\hat{b}_s(\hat{b}^\dagger_{i1}+\hat{b}^\dagger_{i2})+\hat{b}_s^\dagger(\hat{b}_{i1}+\hat{b}_{i2})> \nonumber \\
&&-(e \mathscr{E}_l/2)^2<2\bar{b}_s\bar{b}_s^*+\bar{b}_{i1}\bar{b}^*_{i1}+\bar{b}_{i2}\bar{b}^*_{i2}+\bar{b}_s(\bar{b}^*_{i1}+\bar{b}^*_{i2})+\bar{b}_s^*(\bar{b}_{i1}+\bar{b}_{i2})> \nonumber \\
&=& (e \mathscr{E}_l/2)^2\left[4+<\Delta\hat{b}_s(\Delta\hat{b}^\dagger_{i1}+\Delta\hat{b}^\dagger_{i2})+\Delta\hat{b}_s^\dagger(\Delta\hat{b}_{i1}+\Delta\hat{b}_{i2})>\right] \nonumber \\
&\equiv& (e \mathscr{E}_l)^2\left[1+C(\hat{b}_s,\hat{b}^\dagger_{i1}+\hat{b}^\dagger_{i2})/2\right],
\end{eqnarray}
\end{widetext}
wherein $\hat{b}_s\equiv\hat{a}_se^{i\phi_s}$, $\hat{b}_{i1}\equiv\hat{a}_{i1}e^{i\phi_{i1}}$ and $\hat{b}_{i2}\equiv\hat{a}_{i2}e^{i\phi_{i2}}$. We dropped all the phase-dependent terms after statistical averaging. In the last step, we defined a correlation function $C(\hat{b}_s,\hat{b}^\dagger_{i1}+\hat{b}^\dagger_{i2})\equiv<\Delta\hat{b}_s(\Delta\hat{b}^\dagger_{i1}+\Delta\hat{b}^\dagger_{i2})+\Delta\hat{b}_s^\dagger(\Delta\hat{b}_{i1}+\Delta\hat{b}_{i2})>/2$. In the frame of the current theory of heterodyne detection, the way to understand the experimental results could be as follows: The use of bichromatic local oscillator in the heterodyne detection introduced an extra correlation between light at the two outputs of the beamsplitter, and it was this correlation, which is represented by the function $C(\hat{b}_s,\hat{b}^\dagger_{i1}+\hat{b}^\dagger_{i2})$ in Eq. (\ref{eq:pnn3}), that cancelled the extra quantum noise of the image sideband vacuum modes. An obvious problem with this explanation is origin of this correlation function, which has never been predicted in theory. Of course, there may be other ways to interpret the experimental results. For example, one may develop a theory to explain the experiment by calculating the power spectral density of the photocurrent fluctuations as a Fourier transform of certain two-time auto-correlation function \cite{ou1987,mandel1995}, which is definitely out of the scope of the current work. In any case, revelation of the mechanism for the absence of the image vacuum noise in the observation will lead to a deeper understanding of the origin of the quantum noise in optical detection.

\section{conclusions}
We have studied the quantum noise performance of a phase-sensitive heterodyne detector with a bichromatic local oscillator. We have first addressed a theoretical dilemma inherent to the existing theory of detection about the quantum noise of the phase-sensitive device. Then we have reported on an experiment on the heterodyne detector and the results show that the studied device is noise free at the quantum level. A deeper understanding of the origin of the quantum noise in optical detection will be gained if revealed is the mechanism for the absence of the image sideband vacuum noise in the observation.

\begin{acknowledgments}
This work was supported by the National Natural Science Foundation of China (grant No. 11174094). 
\end{acknowledgments}

\bibliography{NoiselessPIA}

\end{document}